\documentclass[twocolumn,trackchanges]{aastex62}
\usepackage{amsmath}

\shorttitle{HST non-detection of PSR~J2144--3933}
\shortauthors{S. Guillot et al.}

\newcommand{\hst}{\textit{HST}}
\newcommand{\hstlong}{\textit{Hubble Space Telescope}}
\newcommand{\vlt}{\textit{VLT}}
\newcommand{\vltlong}{\textit{Very Large Telescope}}

\newcommand{\xmmlong}{\textit{XMM-Newton}}

\newcommand{\nh}{\mbox{$N_{\rm H}$}}

\newcommand{\rinfty}{\mbox{$R_{\infty}$}}
\newcommand{\tinfty}{\mbox{$T_{\infty}$}}

\newcommand{\simlt}{\mathrel{\hbox{\rlap{\hbox{\lower4pt\hbox{$\sim$}}}\hbox{$<$}}}}
\newcommand{\simgt}{\mathrel{\hbox{\rlap{\hbox{\lower4pt\hbox{$\sim$}}}\hbox{$>$}}}}

\newcommand{\ee}[1]{\mbox{$10^{#1}$}}
\newcommand{\tee}[1]{\mbox{$\times 10^{#1}$}}
\newcommand{\ud}[2]{\mbox{$^{+ #1}_{- #2}$}}

\newcommand{\unit}[1]{\mbox{$\rm\,#1$}}
\def\deg{\hbox{$^\circ$}}
\def\arcdeg{\hbox{$^\circ$}}

\def\arcsec{\hbox{$^{\prime\prime}$}}

\newcommand{\G}{\mbox{$\,G$}}
\newcommand{\msun}{\mbox{$\,M_\odot$}}

\newcommand{\km}{\hbox{$\,{\rm km}$}}
\newcommand{\K}{\hbox{$\,{\rm K}$}}

\newcommand{\yr}{\mbox{$\,{\rm yr}$}}

\newcommand{\pc}{\mbox{$\,{\rm pc}$}}
\newcommand{\persec}{\mbox{$\,{\rm s^{-1}}$}}

\newcommand{\percmsq}{\mbox{$\,{\rm cm^{-2}}$}}
\newcommand{\percmcube}{\mbox{$\,{\rm cm^{-3}}$}}

\newcommand{\cgslum}{\mbox{$\,{\rm erg\,\persec}$}}

\newcommand{\nJy}{\mbox{$\,{\rm nJy}$}}

\begin{document}

\title{Hubble Space Telescope non-detection of PSR~J2144--3933:\\ the coldest known neutron star\footnote{Based on observations made with the NASA/ESA Hubble Space Telescope, obtained at the Space Telescope Science Institute, which is operated by the Association of Universities for Research in Astronomy, Inc., under NASA contract NAS 5-26555. These observations are associated with program \#13783.}
}

\correspondingauthor{Sebastien Guillot}
\email{sebastien.guillot@irap.omp.eu}

\author[0000-0002-6449-106X]{Sebastien Guillot}
\altaffiliation{CNES Post-doctoral Fellow}
\affil{IRAP, CNRS, 9 avenue du Colonel Roche, BP 44346, F-31028 Toulouse Cedex 4, France}
\affil{Universit\'{e} de Toulouse, CNES, UPS-OMP, F-31028 Toulouse, France.}
\affil{Instituto de Astrof\'{i}sica, Pontificia Universidad Cat\'{o}lica de Chile, Av. Vicu\~{n}a Mackenna 4860, Macul, Santiago, Chile}

\author[0000-0002-7481-5259]{George G. Pavlov}
\affil{Pennsylvania State University, Department of Astronomy \& Astrophysics, 525 Davey Lab., University Park, PA 16802, USA}

\author[0000-0003-4294-9647]{Cristobal Reyes}
\affil{Instituto de Astrof\'{i}sica, Pontificia Universidad Cat\'{o}lica de Chile, Av. Vicu\~{n}a Mackenna 4860, Macul, Santiago, Chile}

\author[0000-0003-4059-6796]{Andreas Reisenegger}
\affil{Instituto de Astrof\'{i}sica, Pontificia Universidad Cat\'{o}lica de Chile, Av. Vicu\~{n}a Mackenna 4860, Macul, Santiago, Chile}

\author{Luis E. Rodriguez}
\affil{Instituto de F\'{i}sica, Pontificia Universidad Cat\'{o}lica de Chile, Av. Vicu\~{n}a Mackenna 4860, Macul, Santiago, Chile}

\author[0000-0002-9282-5207]{Blagoy Rangelov}
\affil{Texas State University, Department of Physics, 601 University Drive, San Marcos, TX 78666, USA}

\author{Oleg Kargaltsev}
\affil{George Washington University, Department of Physics, 725 21st Street, NW, Washington, DC 20052, USA}

\begin{abstract}
We report non-detections of the $\sim 3\times 10^8\,\mathrm{yr}$ old, slow, isolated, rotation-powered pulsar PSR~J2144--3933 in observations with the \hstlong{} in one optical band (F475X) and two far-ultraviolet bands (F125LP and F140LP), yielding upper bounds $F_{\rm F475X}< 22.7\nJy$, $F_{\rm F125LP}< 5.9\nJy$, $F_{\rm F140LP}< 19.5\nJy$, at the pivot wavelengths 4940\,\AA, 1438\,\AA\ and 1528\,\AA, respectively. Assuming a blackbody spectrum, we deduce a conservative upper bound on the surface (unredshifted) temperature of the pulsar of $T<42,000\K$. This makes PSR~J2144--3933 the coldest known neutron star, allowing us to study thermal evolution models of old neutron stars. This temperature is consistent with models with either direct or modified Urca reactions including rotochemical heating, and, considering frictional heating from the motion of neutron vortex lines, it puts an upper bound on the excess angular momentum in the neutron superfluid, $J<10^{44}\,\mathrm{erg\,s}$.

\end{abstract}


\section{Introduction}
\label{sec:intro}

The interior composition of neutron stars (NSs) remains largely unknown fifty years after their discovery, but several observational opportunities exist to test models of their interiors. For example, measurements of their mass \citep[e.g.,][]{demorest10} and their radius \citep[e.g.,][]{guillot13,guillot16b,steiner18} provide information on the dense matter in the core of NSs. Other internal microphysics properties (e.g., superfluid energy gaps, the proton fraction, possible ``pasta'' phases) can be probed by studying the evolution of the NS surface temperature on different timescales (for comprehensive reviews, see \citealt{yakovlev04,potekhin15}).

To study the long-term cooling of NSs, one can compare thermal evolution models to the (relatively small) sample of NSs with measured surface temperatures. A general property of passive cooling models is that they predict a rather sharp drop in the temperature once the emission of neutrinos from the core becomes negligible compared to that of thermal photons from the surface, leading to surface temperatures below $\ee{4}\K$ after $\ee{7}\yr$ \citep{yakovlev04,potekhin15}.

Various internal reheating mechanisms had been proposed, but largely ignored until \hstlong\ (\hst) observations showed the nearest millisecond pulsar (MSP), PSR~J0437--4715, a Gyr-old object, to emit far ultraviolet (FUV) radiation consistent with thermal emission from a NS surface at temperature $\sim\ee{5}\K$ \citep{kargaltsev04,durant12}. \citet{gonzalez10} showed that, among the previously proposed heating mechanisms, only two could possibly account for such a high temperature, namely \emph{rotochemical heating} and  \emph{vortex friction}, both driven by the slow decrease in the rotation rate of the star. The former is caused by the decreasing centrifugal force, which compresses the stellar matter, causing a chemical imbalance that drives ``Urca reactions'' in the NS core. These reactions slowly convert different particle species into each other (particularly neutron beta decay into a proton, an electron or muon, and the corresponding anti-neutrino), releasing energy in the form of neutrinos and excess heat \citep{reisenegger95,fernandez05,petrovich10}. In the latter mechanism, the superfluid neutron vortices move out through the crustal lattice as the NS spins down, dissipating energy through friction \citep{alpar84,shibazaki89,larson99}.  

The measurement of the surface temperature of several old NSs would permit testing these mechanisms and characterizing them accurately.  However, at these temperatures ($\sim\ee{5}\K$), the peak of the emission lies in the extreme ultraviolet, where it is strongly absorbed by interstellar atomic hydrogen, making it essentially undetectable.  In X-ray observations, the surface emission can be dominated by the emission from the hot polar caps ($\sim\ee{6}\K$, likely heated by return currents of the magnetosphere, \citealt{arons81,harding01,harding02}).  Disentangling this from the rest of the surface may require broad-band X-ray spectral analysis \citep[see][for the analyses of PSR~J0437--4715]{durant12,guillot16a}. However, in the FUV band, the Rayleigh-Jeans (RJ) tail of the $\sim\ee{5}\K$ surface emission may be dominant and detectable by the \hst.

We have obtained optical and FUV images of three old pulsars (two classical [slow] pulsars and one MSP) to detect their thermal emission RJ tail and measure their surface temperatures. 
For the MSP PSR~J2124--3358, we measured a surface temperature in the range 50,000--260,000\K{}, which accounts for the poor knowledge of the extinction, the uncertain distance, and the possibility of non-thermal emission \citep{rangelov17}. For the classical pulsar PSR~B0950+08, we obtained a surface temperature in the range 130,000--250,000\K{} \citep{pavlov17}, by combining our \hst{} observations to archival \hst{} and \vltlong{} (\vlt) data. Although the distance to PSR~B0950+08 is accurately known, the extinction and the presence of non-thermal emission contribute to the uncertainties of the temperature measurement.

Here, we present the observations of the third pulsar of our \hst{} program, PSR~J2144--3933 (J2144, hereafter). This isolated pulsar is one of the slowest among rotation-powered pulsars, with spin period $P=8.51\,\mathrm{s}$.  Its period and period derivative ($\dot{P}=4.05\tee{-16}\unit{s\persec}$, \citealt{young99}, corrected for the Shklovskii effect; \citealt{shklovskii70}, see also \citealt{deller09,tiengo11}), correspond to a surface dipolar magnetic field $B=1.9\tee{12}\G$, which place J2144 close to (or beyond) the death line of pulsars in the $P$--$\dot{P}$ diagram. These spin properties imply a very low rate of rotational energy loss, $\dot{E}= 2.6\tee{28}\cgslum$, and one of the largest characteristic ages measured for non-recycled pulsars, $\tau\equiv P/[2\dot P]=333\unit{Myr}$. Located at a relatively small and accurately measured distance, $d=165\ud{27}{14}\pc$ \citep{deller09}, or $d=172\ud{20}{15}\pc$, after correction for the Lutz-Kelker bias \citep{lutz73,verbiest10}, J2144 is an ideal target to search for the thermal emission from its surface and constrain its temperature.  Nonetheless, J2144 has never been detected outside of the radio band.  Optical observations with the \vlt{} and X-ray observations with \xmmlong{} placed upper limits on the surface temperature of $T< 2.3\tee{5}\K$, when assuming that the emission comes from the entire surface \citep{tiengo11}.  These authors also reported non-detections with the \emph{Extreme Ultraviolet Explorer} (EUVE; \citealt{bowyer91}) and with the \emph{Galaxy Evolution Explorer} (GALEX; \citealt{martin05}).

Here, we discuss the analysis of the deep optical and FUV observations of J2144 performed by \hst, and present the upper bounds placed on the surface temperature of this old isolated pulsar. We present the data processing, analysis, and results in Section~\ref{sec:obs}, \ref{sec:analysis}, and \ref{sec:results}, respectively. In Section~\ref{sec:discussion}, we discuss the implications of this pulsar's surface temperature upper bound in the context of reheating mechanisms of old NSs.

\begin{deluxetable}{ccccc}[t]
\tablecaption{\hst{} observations of PSR~2144--3933 \label{tab:obs}}
\tablecolumns{5}
\tablewidth{0pt}
\tablehead{
\colhead{Date (MJD)} & \colhead{Instrument} & \colhead{Filter} & \colhead{Exposure (s)}
}
\startdata
57118 & WFC3/UVIS & F475X & 2552 \\
57211 & ACS/SBC   & F125LP & 3695 \\
57211 & ACS/SBC   & F140LP & 1233 \\
57224 & ACS/SBC   & F125LP & 3695 \\
57224 & ACS/SBC   & F140LP & 1233 \\
\enddata
\end{deluxetable}

\section{Observations and data processing}
\label{sec:obs}

Our \hst{} program (\#13783, PI: Pavlov) for J2144 consisted of five orbits in the optical and FUV bands (Table~\ref{tab:obs}). For the former, we have used the Wide Field Camera (WFC3), with the extremely wide F475X filter in the ultraviolet-visible (UVIS) channel. For these exposures, the pulsar position was placed close to the detector readout ($\sim 15\arcsec$ from a corner) to limit the contaminating effects of charge transfer efficiency losses (which reduces the sensitivity for faint objects). The FUV exposures were obtained with the Solar Blind Channel (SBC) detector of the Advanced Camera for Surveys (ACS) with the F125LP and F140LP long-pass filters. For these exposures, the pulsar position was centered on the detector. To limit the effect of strong geocoronal emission lines ($\sim 1400$\AA), we alternatively used the F125LP filter in the Earth's shadow and the F140LP filter outside the Earth shadow. For comparison of the throughput of the filters, see Figure 1 of \cite{rangelov17}.

The data were downloaded from the Mikulski Archive for Space Telescopes\footnote{\url{http://archive.stsci.edu/}} (MAST) and processed with the \emph{DrizzlePac} package (version 2.1.13, provided in the \emph{Astroconda} PYTHON distribution). For the F475X data, the calibrated individual images ({\tt flt}) were first co-aligned using the {\tt Tweakreg} task, where we adjusted the parameters to minimize the residuals of the frame-to-frame point source offsets, indicating the optimal alignment of the frames.  For the FUV data, not enough sources ($<5$) were detected to perform the coalignment with {\tt Tweakreg}. However, visual inspection of the available {\tt flt} files confirms that their World Coordinate System (WCS) appear sufficiently well coaligned. Following this, the standard {\tt AstroDrizzle} task was executed on the sets of frames from each filter to correct for geometric distortion, remove cosmic rays, subtract the sky (if applicable), and combine the frames into a single image.

\begin{figure}
\centering
\includegraphics[width=\columnwidth]{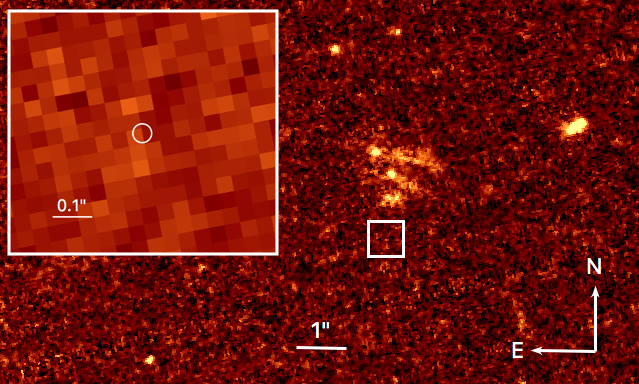} 
\caption{UVIS F475X image of the field around the position of pulsar PSR~J2144--3993. In the inset image, the ellipse shows the 2$\sigma$ uncertainty around the position listed in Table~\ref{tab:pos}. The pulsar is not detected in this image.}
\label{fig:F475X}
\end{figure}

\begin{deluxetable*}{ccllcc}[t]
\tablecaption{Positions of PSR~2144--3933 at different epochs\label{tab:pos}}
\tablecolumns{6}
\tablewidth{0pt}
\tablehead{
\colhead{Epoch (MJD)} & \colhead{Data} &\colhead{RA ($\alpha$)} & \colhead{DEC ($\delta$)} & \colhead{$\mu_{\alpha}$ (mas/yr)} & \colhead{$\mu_{\delta}$ (mas/yr)}
}
\startdata
54100 & VLBI & 21:44:12.06040(5) & $-$39:33:56.8850(3) &	$-$57.89(88) &	$-$155.90(54)\\ 
57118 & UVIS & 21:44:12.019(1)  & $-$39:33:58.174(12) & -- & --\\
57211 & SBC  & 21:44:12.018(43) & $-$39:33:58.214(500) & -- & --\\
57224 & SBC	 & 21:44:12.018(43) & $-$39:33:58.219(500) & -- & --\\
\enddata
\tablecomments{The numbers in parentheses represent the uncertainties in the last digit(s), with all sources of uncertainties included (see Sections~\ref{sec:obs} and \ref{sec:analysis} for details).  Since the SBC images are stacked (for each of the two filters) and aligned with respect to the first SBC image as reference, we used the MJD 57211 position.}
\end{deluxetable*}

\begin{figure*}[t]
\centering
\includegraphics[width=0.9\textwidth]{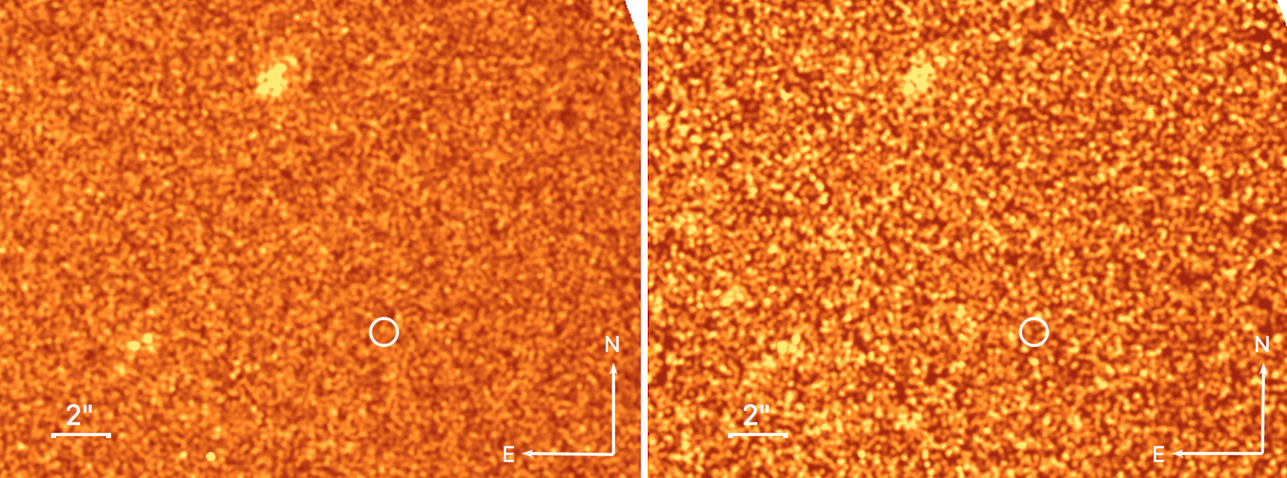} 
\caption{F125LP (left) and F140LP (right)images obtained around the position of pulsar PSR~J2144--3993. These images have been manually aligned to the F475X image (Figure~\ref{fig:F475X}) using the two sources detected to the north and east of the pulsar position. The circle size represents the estimated residual uncertainty of $0\farcs5$. The pulsar is not detected in these images.}
\label{fig:FUV}
\end{figure*}

The astrometric solution of the merged F475X image was then improved using the GAIA DR1 Catalog \citep{gaia16a,gaia16b}.  While 15 cataloged objects fall within the field of view of the F475X image, we excluded a few saturated or possibly extended objects, resulting in 11 objects used for astrometric correction. Using \emph{IRAF}'s task \emph{CCMAP}, we obtained an image aligned to the GAIA catalog, with root-mean-square (rms) residuals of 0\farcs0026 in right ascension (RA) and 0\farcs0067 in declination (DEC).  We discuss below how these residuals are added to other sources of uncertainties affecting the pulsar position. Note that no proper motion data were available for these catalog stars, but because the epochs of observations of the GAIA catalog (between MJD 56863 and MJD 57281) and of our \hst{} image do not differ by more than 8.4 months, the effects of proper motions are likely to be negligible. 

The FUV ACS/SBC images contain no cataloged sources, and only two sources with counterparts in the F475X image (one of which is an extended source). Comparison of the positions of these two sources in the F475X and FUV images reveals an offset of $\sim 0\farcs4$ in RA and $\sim 0\farcs9$ in DEC. We correct the World Coordinate System of the FUV images to align it to the F475X image. We conservatively estimate the residual error of this alignment to be $\sim0\farcs5$.

\section{Analysis}
\label{sec:analysis}
In these processed images, we search for emission from J2144 at its expected position at the time of the \hst{} observations. The pulsar position and proper motion were obtained from Very Long Baseline Interferometry observations \citep{deller09}, with reference epoch MJD 54100. Accounting for the source proper motion, we calculated the pulsar coordinates for each of the epochs of our observations. These positions are summarized in Table~\ref{tab:pos}.  The uncertainties on these calculated positions include the rms residuals from the astrometric correction in RA and DEC (see Section~\ref{sec:obs}), the uncertainties due to the propagation of the pulsar proper motion uncertainties, and the systematic uncertainties of the GAIA catalog stars used for the astrometric correction (0\farcs01, \citealt{lindegren16}). These various sources uncertainties are added in quadrature and reported in  Table~\ref{tab:pos}. For the two FUV images, the estimated $\sim0\farcs5$ residual error from the manual alignment dominates over the other sources of uncertainties.

Figure~\ref{fig:F475X} shows the F475X image and the expected position of the pulsar. In the inset, the ellipse has a size of 0\farcs023 in RA and 0\farcs024 in DEC, corresponding to twice the astrometric uncertainties reported in Table~\ref{tab:pos} and Section~\ref{sec:obs}. Visual inspection does not reveal any source at the calculated pulsar position, nor does the source detection algorithm (Python package {\tt photutils}, based on the {\tt daofind IRAF} package), with a threshold of $3\sigma$.  Similarly, there is no evidence of a source at the pulsar position in the two FUV images (F125LP and F140LP, Figure~\ref{fig:FUV}).

\begin{deluxetable*}{lcccccccccc}[t]
\tablecaption{Count rates and mean flux density upper bounds of \mbox{PSR~J2144--3933}. \label{tab:flux}}
\tablecolumns{10}
\tablewidth{0pt}
\tablehead{
\colhead{Image} & \colhead{$\lambda_{\rm piv}$} &\colhead{$t_{\rm exp}$} & \colhead{$r_{\rm extr}$} & \colhead{$\phi_{\rm E}$} &\colhead{$C_{\rm pos}$ } &\colhead{$\bar{C}_{\rm bkg}$} &\colhead{$\sigma_{\rm bkg}$} & \colhead{$C_{\rm ub}$ } & \colhead{${\cal P}_\nu$} & \colhead{$f_{\nu,\,{\rm ub}}$} \\
\colhead{(filter)} & \colhead{(\AA)}& \colhead{(s)} & \colhead{($''$)} & \colhead{(\%)}& \colhead{(cts/ks)} & \colhead{(cts/ks)} & \colhead{(cts/ks)} & \colhead{(cts/ks)} & \colhead{(nJy ks/cts)} & \colhead{(\nJy)}
}
\startdata
\hline
F475X  & 4940 & 2552 & 0.126 & 75 & $-$14.0 & $-$35.7 & 37.8 & 135.1 & 0.126 & 22.7 \\
F125LP & 1438 & 7390 & 0.15  & 57 & 3.0  & 2.1 & 0.64 & 2.82 & 1.19  & 5.9 \\
F140LP & 1528 & 2466 & 0.187 & 62 & 4.9  & 3.6 & 1.47 & 5.71 & 2.12  & 19.5 \\
\enddata
\tablenotetext{}{All count rates are provided for the extraction radii $r_{\rm extr}$.
$C_{\rm pos}$ is the count rate at the pulsar position. $\bar{C}_{\rm bkg}$ and $\sigma_{\rm bkg}$ are the mean and standard deviation of the background. $C_{\rm ub}$ is the count rate upper bound given by $C_{\rm ub} = C_{\rm pos} - \bar{C}_{\rm bkg} + n \sigma_{\rm bkg}$, for $n=3$ (see Section~\ref{sec:analysis}). $f_{\nu,\,{\rm ub}}$ is the aperture-corrected upper bound, using the encircled energy fraction $\phi_{\rm E}$ on the mean flux density for the conversion factors ${\cal P}_\nu$ and the pivot wavelength $\lambda_{\rm piv}$ of each filter.}
\end{deluxetable*} 

We therefore calculate bounds on the minimum detectable fluxes at the pulsar position for each of the three \hst{} images. To do so, we define the upper bounds $C_{\rm ub}$ on the count rate as $C_{\rm ub} = C_{\rm pos} - \bar{C}_{\rm bkg} + n \sigma_{\rm bkg}$ (see \citealt{kashyap10}), where $C_{\rm pos}$ is the measured count rate at the position of the pulsar, $\bar{C}_{\rm bkg}$ and $\sigma_{\rm bkg}$ are the mean and standard deviation of the background count rate, and $n$ determines the significance of the upper bound we calculate. We use the term \emph{upper bound} rather than \emph{upper limit}, according to the discussion of \cite{kashyap10}.

The sizes of the extraction regions used to measure $C_{\rm pos}$ were chosen by identifying on each image the extraction radius that maximizes the signal-to-noise ratio (S/N), using a nearby detected faint point source. For the F475X image, the optimal extraction radius is $r_{\rm extr}=0\farcs126$ (3.14 pixels), corresponding to 75\% enclosed energy fraction\footnote{The enclosed energy fraction for WFC3/UVIS is available at \url{http://www.stsci.edu/hst/wfc3/analysis/uvis\_ee/\#Tables}.} $\phi_{\rm E}$, and the extraction circle to measure $C_{\rm pos}$ is centered on the calculated position (see Table~\ref{tab:pos}), since the error on the latter is smaller than the extraction region.  In the FUV band with the ACS/SBC instrument, the optimal radius is $r_{\rm extr}=0\farcs15$ (6 pixels; $\phi_{\rm E}=57\%$) for the F125LP filter, and $r_{\rm extr}=0\farcs187$ (7.5 pixels; $\phi_{\rm E}=62\%$) for the F140LP filter\footnote{The enclosed energy fraction for ACS/SBC is available at \url{http://www.stsci.edu/hst/acs/documents/isrs/isr1605.pdf}.}. In both FUV images, the error circle ($0\farcs5$ radius) is larger than the optimal extraction circles.  Therefore, we measure $C_{\rm pos}$ from the extraction circle placed at the position within the error circle that maximizes the $C_{\rm pos}$ value.

We estimate $\bar{C}_{\rm bkg}$ and $\sigma_{\rm bkg}$ from a collection of background circular regions of size $r_{\rm extr}$ (see above, and Table~\ref{tab:flux}). The F475X image exhibits a large-scale background gradient. Therefore, $\sim 100$ non-overlapping background regions are randomly selected along a line of constant background, and passing through the pulsar position. Using {\tt photutils}, we obtain the sum of counts in each of these regions and calculate the mean $\bar{C}_{\rm bkg}$ and standard deviation $\sigma_{\rm bkg}$  needed to obtain $C_{\rm ub}$. Outliers (4 standard deviations in excess of the average count rate) of these randomly selected regions are excluded as they may correspond to the location of real sources. This procedure is repeated 200 times, each with $\sim 100$ randomly generated regions, to verify that $\bar{C}_{\rm bkg}$ and $\sigma_{\rm bkg}$ do not vary significantly between trials. Finally, we also perform some trials by moving the ``background line'' end points by about $\pm 1\arcsec$, but still passing through the pulsar position, to be convinced that the choice of this line does not significantly affect $\bar{C}_{\rm bkg}$ and $\sigma_{\rm bkg}$. For the F125LP and F140LP images, which do not show a background gradient, the 100 background regions are chosen within about $\pm 15\arcsec$ of the pulsar position. The rest of the procedure to calculate $\sigma_{\rm bkg}$ is the same as for the F475X image. 

In the remainder of this work, we choose $n=3$ and therefore report $3\sigma$ upper bounds on the count rate and flux. Section~\ref{sec:results} describes how these count rates are converted into flux upper bounds in the three bands and into a temperature upper bound for J2144.

Finally, we note that the structure north of the pulsar position was reported as a galaxy with a {\it ``somewhat patchy morphology''} based on observations with the FOcal Reducer/low dispersion Spectrograph (FORS2) at the \vlt{} \citep{tiengo11}. The higher angular resolution provided by our \hst{} image confirms the irregular morphology of this object, likely to be a barred galaxy with asymmetric arms.

\section{Upper bound on the pulsar surface temperature}
\label{sec:results}

Table~\ref{tab:flux} summarizes the upper bounds on the count rates and on the mean flux densities that we obtained following the method described in Section~\ref{sec:analysis}. To convert count rates to mean flux densities, we use the conversion factors ${\cal P}_{\nu}$ extracted from the PHOTFNU keyword value in the header of the F475X data files provided by the \hst{} analysis pipeline; and from \cite{pavlov17}, derived from the PHOTFLAM keyword values, for the F125LP and F140LP images (see values of ${\cal P}_{\nu}$ in Table~\ref{tab:flux}).

\begin{figure}[t]
\centering
\includegraphics[width=\columnwidth]{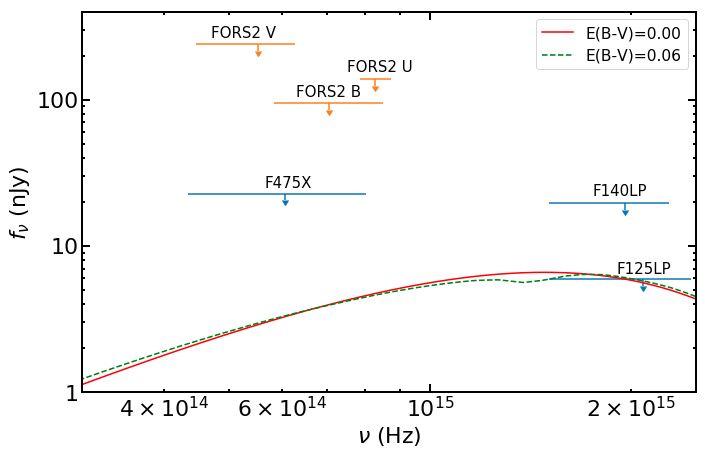} 
\caption{Flux density upper bounds obtained from observations of PSR~J2144--3933 with the \hst{} (blue; results reported in this paper) and the \vlt{} (orange; \citealt{tiengo11}), together with the hottest allowed Planck spectra, for two choices of the reddening $E(B-V)=0.0$ (red solid line) and $E(B-V)=0.06$ (green dashed line), yielding $T=31,950\K$ and $T=36,250\K$, respectively. These are inferred temperatures at the NS surface, corrected for gravitational redshift assuming a NS mass $M=1.4\msun$, coordinate radius $R=11\km$, and distance $d=172\pc$.}
\label{fig:flux}
\end{figure}

Our \hst\ observations provide stringent bounds on the flux densities in the FUV bands, shown in Figure~\ref{fig:flux}. In the optical band, we improved the detection limit by about one order of magnitude over the FORS2 data \citep{tiengo11}. As readily noticeable, our F125LP flux density upper bound dominates the constraints on the thermal emission. 

To estimate the maximum surface temperature allowed by these flux upper bounds, we model the pulsar surface emission with a Planck function $B_\nu(T_\infty)=(2h\nu^3/c^2) [\exp(h\nu/kT_\infty) -1]^{-1}$, including extinction, such that
\begin{equation}
f_\nu = \left(\frac{\rinfty}{d}\right)^2\pi B_\nu(\tinfty)\tee{-0.4 A_\nu}.
\label{eq:flux}
\end{equation}
Here \tinfty\ is the temperature inferred from the spectral energy distribution observed by a distant observer, given in terms of the blackbody (BB) temperature $T$ measured by an observer at the surface as $\tinfty=T/\left(1+z\right)=T\left(1-2 GM/R c^{2}\right)^{1/2}$. $\rinfty$ is the corresponding ``apparent radius'' inferred by the distant observer, defined as $\rinfty=R\left(1+z\right)=R\left(1-2 GM/R c^{2}\right)^{-1/2}$. In Figure 3, we assumed a distance $d=172\pc$ \citep{verbiest10}, as well as typical values for the NS coordinate radius, $R=11\km$, and mass, $M=1.4\msun$, which yield $\rinfty = 13.9\km$.

For the extinction $A_{\nu}$ in Equation~\ref{eq:flux}, we used the extinction curves $A_{\lambda}/A_{V}$ of \cite{clayton03}, extrapolated for values between $2.3\tee{15}$ and $2.5\tee{15}\unit{Hz}$ since they are not defined above $2.3\tee{15}\unit{Hz}$ (see their Figure 5). The amount of extinction/reddening is however rather uncertain. The radio dispersion measure of J2144 is ${\rm DM}=3.35\unit{pc\percmcube}$ \citep{damico98}, which corresponds to a neutral hydrogen column density $\nh=1.0\tee{20}\percmsq$ using the relation of \cite{he13}. This value of \nh{} is not too dissimilar from the estimate obtained from a neutral hydrogen map\footnote{Using the HEASARC tool {\tt nH Column Density} at \url{heasarc.gsfc.nasa.gov/Tools/generaltools.html}.}: $\nh=2.0\tee{20}\percmsq$ \citep{kalberla05} at the Galactic coordinates of J2144: $l = 2.79\deg$, $b=-49.47\deg$. In addition, using the relation between \nh{} and $E(B-V)$ of \cite{guver09}, we find $E(B-V)=0.015$ or $0.03$, for the two values of \nh{} mentioned above.  Furthermore, Galactic dust and extinction maps\footnote{Available at NASA/IPAC, \url{irsa.ipac.caltech.edu/applications/DUST/}} give $E(B-V)=0.028-0.033$ in the direction of J2144 \citep{schlegel98,schlafly11}, consistent with the other values. We note that neutral H maps and Galactic dust and extinction maps provide estimates of the total integrated line-of-sight values, and therefore represent an upper limit to the \nh\ for a nearby object like J2144. Finally, the 3D extinction map of \cite{lallement14} gives $E(B-V)=0.016\pm0.016$ for a distance of $d=175\pc$. Given the uncertainties in the relations mentioned above, we use a very conservative range $E(B-V)=0.00-0.06$ for our calculations and show the two extrema in Figure~\ref{fig:flux}. In Equation~\ref{eq:flux} and throughout this work, the assumed visual extinction to reddening ratio is $A_{V} / E(B-V) = 3.1$ \citep{schultz75}.

\begin{figure}[t]
\centering
\includegraphics[width=\columnwidth]{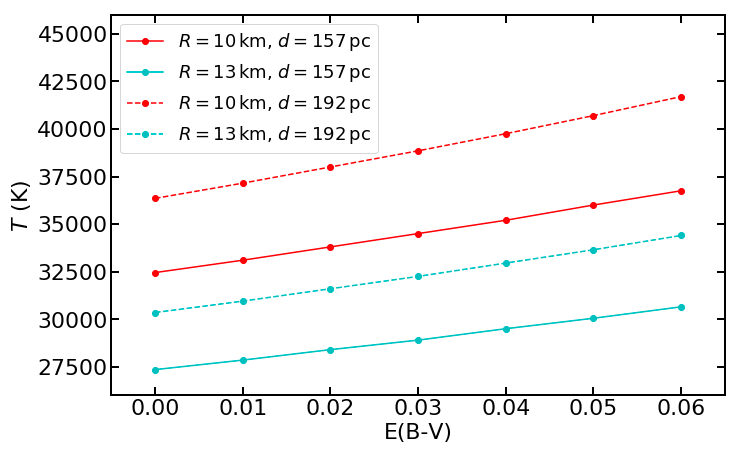} 
\caption{Upper bounds on the unredshifted BB temperature imposed by the F125LP flux upper bound as a function of extinction $E(B-V)$. The solid and dashed lines were obtained for 157\pc{} and 192\pc, the lower and upper values of the 1$\sigma$ confidence interval of the distance \citep{verbiest10}, respectively. The red and cyan colors correspond to NS radii of 10 and 13\km, respectively. The maximum surface BB temperature allowed by the chosen parameters is 42,000\K. These curves were obtained assuming a NS mass of 1.4\msun.}
\label{fig:flux2}
\end{figure}

Figure~\ref{fig:flux} shows the observational upper bounds on the flux in various bands, together with the unabsorbed and absorbed Planck functions for $E(B-V)=0$ and $0.06$, in each case at the temperature yielding our upper bound on the F125LP flux, $T=31,950\K$ and $T=36,250\K$, respectively. It can be seen that there are very minor differences between the curves. Given the upper bounds on the F125LP flux density, a larger assumed extinction permits accommodating a larger BB temperature. 

A non-thermal component (presumably of magnetospheric origin) is often  observed in the optical-UV spectra of pulsars \citep[e.g.][]{kargaltsev07,mignani11}. Considering the possibility of non-thermal emission by adding a power-law component would decrease the contribution of the thermal component to the total source flux, thus also decreasing the BB temperature upper bound deduced from our analysis. In this sense, the temperature upper bounds we obtain by assuming a single thermal component are the most conservative estimates.
Similarly, considering the contribution from the Rayleigh-Jeans tail of the thermal emission of small (tens to hundreds of meters emitting area), but much hotter polar caps ($\sim \ee{6}\K$) would also have the effect of reducing the maximum temperature allowed for the bulk surface emission. We note that the polar cap sizes and temperatures constrained by the X-ray limits of \cite{tiengo11} would contribute $\sim 20\%$ (for the 500-m $5\tee{5}\K$ polar cap) or $<0.1\%$ (for the 10-m $1.9\tee{6}\K$ polar cap).

We also explore the different temperature bounds given by our data for a range of possible distances and NS radii. Indeed, the distance to J2144 is measured from radio parallax with non-negligible uncertainties: $d=172\ud{20}{15}\pc$ \citep{verbiest10}. Similarly, the NS radius is also uncertain since the equation of state of dense matter is not yet known; typical values are expected in the 10--13\km\ range \citep[e.g.,][]{guillot16b,ozel16a,steiner18,abbott17}. This exploration of the parameter space is summarized in Figure~\ref{fig:flux2}, where the surface temperature upper bound is shown as a function of extinction $E(B-V)$ for various distances and radii. As expected from Equation~\ref{eq:flux}, increasing the source distance means that a hotter NS is needed to attain our measured flux upper bound, whereas changing the NS radius has the opposite effect. Overall, the maximum temperature allowed by the maximum possible distance, the minimum acceptable NS radius, and the largest $E(B-V)$ considered is $T=42,000\K$.

Assuming purely non-thermal emission (modeled with a power-law model), we find that the F475X and F125LP constrain the optical non-thermal luminosity to $L_{\rm opt}\sim 5\tee{26}\cgslum$ at a distance of 172\pc; using the range 4000--9000\,\AA. With the very low spin-down power of J2144, $\dot{E}=2.6\tee{28}\cgslum$, we obtain an upper bound on the magnetospheric optical efficiency $\eta_{\rm opt}=L_{\rm opt}/\dot{E} < 0.02$, which is not an interesting constraint as typical values for old pulsars are $\eta_{\rm opt}\sim \ee{-5}$ \citep{zavlin04}.

\section{Discussion and conclusions}
\label{sec:discussion}

Our deep \hst{} observations of the pulsar J2144 in the optical and FUV bands resulted in upper bounds on the flux densities. Using conservative assumptions and a range of possible NS radii and distances, we find that the surface temperature of this old NS cannot be larger than 42,000\K{}, making it the coldest NS known. This bound is driven exclusively by the non-detection in the F125LP FUV band, the most restrictive of our measurements. 

The obtained upper bound also constrains the properties of reheating mechanisms proposed for old, non-recycled, pulsars, especially when combined with our recent temperature measurement for PSR~B0950+08 (hereafter ``B0950''; \citealt{pavlov17}), which requires some heating mechanisms to explain its substantially higher surface temperature, 130,000--250,000\K.

\begin{figure}[t]
\centering
\includegraphics[width=\columnwidth]{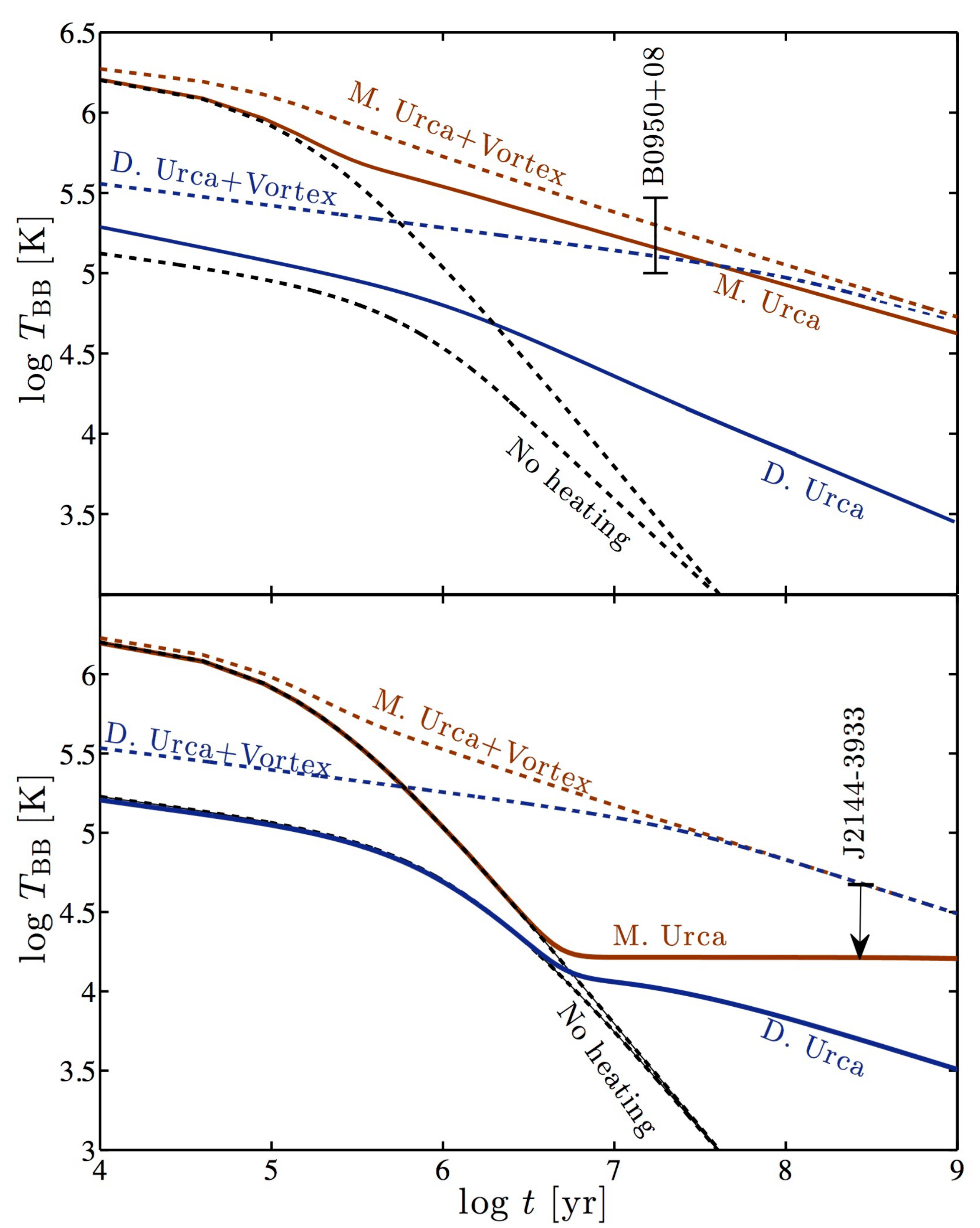} 
\caption{Thermal evolution models for PSR~B0950+08 (upper panel) and PSR~J2144-3933 (lower panel), constrained by the temperature measurement of \citet{pavlov17} and the upper bound obtained in this work. The model curves correspond to different scenarios involving direct or modified Urca processes, with or without vortex friction in addition to rotochemical heating (see \citealt{gonzalez10} for a discussion), assuming $M=1.4\msun$, $R=10\km$, a constant magnetic field (as measured for each pulsar), a fast initial spin period, $P_{0}=1\unit{ms}$, and a high initial temperature, $T_{0}=\ee{11}\K$. Solid curves show the evolution with rotochemical heating, either with only modified Urca reactions (red brown) or with direct Urca reactions (blue). The dashed curves include frictional heating due to superfluid vortex creep (with excess angular momentum $J=\ee{44}\unit{erg\,s}$) in addition to rotochemical heating, again with only modified Urca reactions (red brown) or with direct Urca reactions (blue). In all cases, the possible effect of Cooper pairing gaps on the Urca reactions is ignored. We use the A18+$\delta$ $v$+ UIX* EOS \citep{akmal98} for the models with modified Urca reactions and the BPAL32 EOS \citep{prakash88} for those with direct Urca reactions.}
\label{fig:cooling}
\end{figure}

Figure~\ref{fig:cooling} shows a selection of thermal evolution models for B0950 (upper panel) and J2144 (lower panel) under different scenarios, considering rotochemical heating and vortex friction. Both processes depend on the spin-down history of the NS, which is unknown in detail, so we adopt a simple model of magnetic spin-down with a constant dipole strength inferred from the present spin-down parameters of each pulsar ($B^2\propto P\dot P$), which is $B=2.4\times 10^{11}\,\mathrm{G}$ for B0950 and $1.9\times 10^{12}\,\mathrm{G}$ for J2144, so the latter pulsar spins down much more quickly, making its temperature at late times lower, particularly for rotochemical heating. The spin-down history still depends on one free parameter, namely the rotation period at birth, which we chose very short, $P_0=1\,\mathrm{ms}$, for both pulsars, leading to maximal rotochemical heating \citep{gonzalez10}. For simplicity, we also ignore the effects of Cooper pairing, which would introduce a dependence on the largely unknown energy gaps of neutrons and protons.

It can be seen that the rotochemical heating mechanism is consistent with the upper bound for J2144, as it predicts lower temperatures, $T \sim \ee{3.6}\K$ and $T \sim \ee{4.2}\K$ for direct and modified Urca reactions, respectively, at the time the model star reaches the current spin parameters $P$ and $\dot P$ of this pulsar (corresponding to its inferred characteristic age $\tau \sim \ee{8.4}\yr$), even for the very fast initial rotation assumed in our models, $P_0$ = 1 ms. For longer (and thus likely more realistic) initial spin periods, the chemical imbalance caused by spin-down will be smaller, therefore even cooler temperatures would be obtained \citep{gonzalez10}. However, as pointed out before \citep{pavlov17}, this mechanism alone (and without large Cooper pairing gaps) can only explain the higher temperature of B0950 if the initial rotation is extremely fast ($P_0<10\,\mathrm{ms}$) and no direct Urca reactions are allowed.

As also seen in Figure~\ref{fig:cooling}, the addition of vortex friction causes the thermal evolution curves with direct and modified Urca reactions to converge beyond $\sim \ee{7}\yr$, because photon emission becomes the main cooling mechanism compensating for the frictional heating. With this mechanism, the heat released is proportional to the excess angular momentum $J$ of the superfluid component. The curves shown correspond to the maximum value of $J$ permitted by our temperature upper bound for J2144, $J=\ee{44}\unit{erg\,s}$, which is also compatible with the measurement for B0950.

Overall, the model constraints derived from our upper bound on the surface temperature of J2144 are consistent  with  those  obtained for B0950. A more extensive study of the parameter space of possible reheating mechanisms including the possible presence of Cooper pairing gaps and considering all the available observations and constraints on thermal emission from old NSs will be presented in a forthcoming paper (Rodr{\'i}guez et al., in preparation).

It is important to keep in mind that some caveats remain; however they should not change the main conclusions of this work:
\begin{itemize}
\item We have used Planck functions to fit the thermal emission of J2144 and those of the other two pulsars we studied, PSR~J2124--3358 \citep{rangelov17} and PSR~B0950+08 \citep{pavlov17}.  However, it has been suggested that NSs may possess a single-composition atmosphere; usually, hydrogen or helium \citep[e.g.,][]{zavlin96,shibanov92}, although carbon is also possible \citep{ho09,klochkov13}. Iron atmosphere models have also been proposed \citep{rajagopal96,zavlin96}. Realistic models of NS atmospheres ought to be used when fitting the spectra of NSs. For J2144, one would need to use low-temperature NS atmosphere models, considering partial ionizations and possibly plasma effects, which is beyond the scope of this paper. 
Alternatively, the possibility exists that no atmosphere remains at such low effective temperatures and strong magnetic fields. If this is the case, the emission from the solid (or perhaps liquid) NS surface would have to be considered.
\item In the thermal evolution modeling, we assumed magnetic dipole spin-down with a constant dipole moment (inferring its value from the present $P$ and $\dot P$) and thus a braking index $n=3$, and with a very fast initial spin period, $P_0=1\,\mathrm{ms}$. For MSPs, it has been shown that the rotochemical heating process should have reached an equilibrium state in which the chemical imbalance and temperature are set by the current spin parameters $P$ and $\dot P$ \citep{reisenegger95,fernandez05}. However, somewhat younger pulsars like J2144 could still be radiating away the chemical energy accumulated during its earlier spin-down, therefore its current temperature depends on its initial spin period \citep{gonzalez10} and on possible deviations from the simple spin-down law. This is not the case for vortex friction, in which the balance between heating and cooling occurs much faster than the inferred age of this pulsar, therefore the expected temperature depends only on the current $P$ and $\dot P$ \citep{gonzalez10}.
\end{itemize} 

In summary, while our \hst{} observations of PSR~J2144--3933 did not result in any detection, we significantly improved the constraints over the previously published flux upper bounds. In the optical, our F475X \hst{} upper bound is a factor of at least $\sim4\times$ more constraining than the \vlt{} observations in the $U$, $B$ and $V$ bands.  In the far-UV, we improved the flux upper bound by a factor of $\sim$1000 over the \emph{GALEX} measurements reported in \cite{tiengo11}.  With these stringent bounds, we constrained the surface BB temperature of J2144 to be less than $42,000\K$, compared to the previous upper bound of $\sim 230,000\K$ \citep{tiengo11}, assuming emission from the entire surface of this old pulsar.

\acknowledgments
We thank Andr\'es Jord\'an for useful discussions on statistics and determinations of flux upper bounds. Support for \hst{} program \#13783 was provided by NASA through a grant from the Space Telescope Science Institute, which is operated by the Association of Universities for Research in Astronomy, Inc., under NASA contract NAS 5-26555. S.G. acknowledges the support of the French Centre National d'\'{E}tudes Spatiales (CNES), of the ECOS-CONICYT program (grant \#C16U01), and of the FONDECYT Postdoctoral Project 3150428 in the early phases of this work. The work of C.R., A.R., and L.R. was supported by FONDECYT Regular Project 1150411 and CONICYT project Basal AFB-170002 (CATA).

\vspace{5mm}
\facilities{\hst/WFC3; \hst/ACS}

\software{This research made use of the following softwares and packages: (1) \texttt{astropy}, a community-developed core Python package for Astronomy \citep{astropy13} available at \url{www.astropy.org}; (2)
          \texttt{drizzlepac}, \citep{gonzaga12} available at \url{drizzlepac.stsci.edu}, (3) \texttt{PhotUtils}, an Astropy affiliated package, available at \url{photutils.readthedocs.io}, and (4) \texttt{IRAF}, distributed by the National Optical Astronomy Observatories \citep{tody93}. }

\bibliographystyle{aasjournal}
\bibliography{biblio}

\begin{thebibliography}{}
\expandafter\ifx\csname natexlab\endcsname\relax\def\natexlab#1{#1}\fi
\providecommand{\url}[1]{\href{#1}{#1}}

\bibitem[{{Abbott} {et~al.}(2017){Abbott}, {Abbott}, {Abbott}, {Acernese},
  {Ackley}, {Adams}, {Adams}, {Addesso}, {Adhikari}, {Adya}, \&
  et~al.}]{abbott17}
{Abbott}, B.~P., {Abbott}, R., {Abbott}, T.~D., {et~al.} 2017, Physical Review
  Letters, 119, 161101

\bibitem[{{Akmal} {et~al.}(1998){Akmal}, {Pandharipande}, \&
  {Ravenhall}}]{akmal98}
{Akmal}, A., {Pandharipande}, V.~R., \& {Ravenhall}, D.~G. 1998, \prc, 58, 1804

\bibitem[{{Alpar} {et~al.}(1984){Alpar}, {Pines}, {Anderson}, \&
  {Shaham}}]{alpar84}
{Alpar}, M.~A., {Pines}, D., {Anderson}, P.~W., \& {Shaham}, J. 1984, \apj,
  276, 325

\bibitem[{{Arons}(1981)}]{arons81}
{Arons}, J. 1981, \apj, 248, 1099

\bibitem[{{Astropy Collaboration} {et~al.}(2013){Astropy Collaboration},
  {Robitaille}, {Tollerud}, {Greenfield}, {Droettboom}, {Bray}, {Aldcroft},
  {Davis}, {Ginsburg}, {Price-Whelan}, {Kerzendorf}, {Conley}, {Crighton},
  {Barbary}, {Muna}, {Ferguson}, {Grollier}, {Parikh}, {Nair}, {Unther},
  {Deil}, {Woillez}, {Conseil}, {Kramer}, {Turner}, {Singer}, {Fox}, {Weaver},
  {Zabalza}, {Edwards}, {Azalee Bostroem}, {Burke}, {Casey}, {Crawford},
  {Dencheva}, {Ely}, {Jenness}, {Labrie}, {Lim}, {Pierfederici}, {Pontzen},
  {Ptak}, {Refsdal}, {Servillat}, \& {Streicher}}]{astropy13}
{Astropy Collaboration}, {Robitaille}, T.~P., {Tollerud}, E.~J., {et~al.} 2013,
  \aap, 558, A33

\bibitem[{{Bowyer} \& {Malina}(1991)}]{bowyer91}
{Bowyer}, S., \& {Malina}, R.~F. 1991, Advances in Space Research, 11, 205

\bibitem[{{Clayton} {et~al.}(2003){Clayton}, {Wolff}, {Sofia}, {Gordon}, \&
  {Misselt}}]{clayton03}
{Clayton}, G.~C., {Wolff}, M.~J., {Sofia}, U.~J., {Gordon}, K.~D., \&
  {Misselt}, K.~A. 2003, \apj, 588, 871

\bibitem[{{D'Amico} {et~al.}(1998){D'Amico}, {Stappers}, {Bailes}, {Martin},
  {Bell}, {Lyne}, \& {Manchester}}]{damico98}
{D'Amico}, N., {Stappers}, B.~W., {Bailes}, M., {et~al.} 1998, \mnras, 297, 28

\bibitem[{{Deller} {et~al.}(2009){Deller}, {Tingay}, {Bailes}, \&
  {Reynolds}}]{deller09}
{Deller}, A.~T., {Tingay}, S.~J., {Bailes}, M., \& {Reynolds}, J.~E. 2009,
  \apj, 701, 1243

\bibitem[{{Demorest} {et~al.}(2010){Demorest}, {Pennucci}, {Ransom}, {Roberts},
  \& {Hessels}}]{demorest10}
{Demorest}, P.~B., {Pennucci}, T., {Ransom}, S.~M., {Roberts}, M.~S.~E., \&
  {Hessels}, J.~W.~T. 2010, \nat, 467, 1081

\bibitem[{{Durant} {et~al.}(2012){Durant}, {Kargaltsev}, {Pavlov}, {Kowalski},
  {Posselt}, {van Kerkwijk}, \& {Kaplan}}]{durant12}
{Durant}, M., {Kargaltsev}, O., {Pavlov}, G.~G., {et~al.} 2012, \apj, 746, 6

\bibitem[{{Fern{\'a}ndez} \& {Reisenegger}(2005)}]{fernandez05}
{Fern{\'a}ndez}, R., \& {Reisenegger}, A. 2005, \apj, 625, 291

\bibitem[{{Gaia Collaboration} {et~al.}(2016{\natexlab{a}}){Gaia
  Collaboration}, {Prusti}, {de Bruijne}, {Brown}, {Vallenari}, {Babusiaux},
  {Bailer-Jones}, {Bastian}, {Biermann}, {Evans}, \& et~al.}]{gaia16a}
{Gaia Collaboration}, {Prusti}, T., {de Bruijne}, J.~H.~J., {et~al.}
  2016{\natexlab{a}}, \aap, 595, A1

\bibitem[{{Gaia Collaboration} {et~al.}(2016{\natexlab{b}}){Gaia
  Collaboration}, {Brown}, {Vallenari}, {Prusti}, {de Bruijne}, {Mignard},
  {Drimmel}, {Babusiaux}, {Bailer-Jones}, {Bastian}, \& et~al.}]{gaia16b}
{Gaia Collaboration}, {Brown}, A.~G.~A., {Vallenari}, A., {et~al.}
  2016{\natexlab{b}}, \aap, 595, A2

\bibitem[{{Gonzaga} \& {et al.}(2012)}]{gonzaga12}
{Gonzaga}, S., \& {et al.} 2012, {The DrizzlePac Handbook}

\bibitem[{{Gonzalez} \& {Reisenegger}(2010)}]{gonzalez10}
{Gonzalez}, D., \& {Reisenegger}, A. 2010, \aap, 522, A16

\bibitem[{{Guillot}(2016)}]{guillot16b}
{Guillot}, S. 2016, \memsai, 87, 521

\bibitem[{{Guillot} {et~al.}(2013){Guillot}, {Servillat}, {Webb}, \&
  {Rutledge}}]{guillot13}
{Guillot}, S., {Servillat}, M., {Webb}, N.~A., \& {Rutledge}, R.~E. 2013, \apj,
  772, 7

\bibitem[{{Guillot} {et~al.}(2016){Guillot}, {Kaspi}, {Archibald}, {Bachetti},
  {Flynn}, {Jankowski}, {Bailes}, {Boggs}, {Christensen}, {Craig}, {Hailey},
  {Harrison}, {Stern}, \& {Zhang}}]{guillot16a}
{Guillot}, S., {Kaspi}, V.~M., {Archibald}, R.~F., {et~al.} 2016, \mnras, 463,
  2612

\bibitem[{{G{\"u}ver} \& {{\"O}zel}(2009)}]{guver09}
{G{\"u}ver}, T., \& {{\"O}zel}, F. 2009, \mnras, 400, 2050

\bibitem[{{Harding} \& {Muslimov}(2001)}]{harding01}
{Harding}, A.~K., \& {Muslimov}, A.~G. 2001, \apj, 556, 987

\bibitem[{{Harding} \& {Muslimov}(2002)}]{harding02}
---. 2002, \apj, 568, 862

\bibitem[{{He} {et~al.}(2013){He}, {Ng}, \& {Kaspi}}]{he13}
{He}, C., {Ng}, C.-Y., \& {Kaspi}, V.~M. 2013, \apj, 768, 64

\bibitem[{{Ho} \& {Heinke}(2009)}]{ho09}
{Ho}, W.~C.~G., \& {Heinke}, C.~O. 2009, \nat, 462, 71

\bibitem[{{Kalberla} {et~al.}(2005){Kalberla}, {Burton}, {Hartmann}, {Arnal},
  {Bajaja}, {Morras}, \& {P{\"o}ppel}}]{kalberla05}
{Kalberla}, P.~M.~W., {Burton}, W.~B., {Hartmann}, D., {et~al.} 2005, \aap,
  440, 775

\bibitem[{{Kargaltsev} \& {Pavlov}(2007)}]{kargaltsev07}
{Kargaltsev}, O., \& {Pavlov}, G. 2007, \apss, 308, 287

\bibitem[{{Kargaltsev} {et~al.}(2004){Kargaltsev}, {Pavlov}, \&
  {Romani}}]{kargaltsev04}
{Kargaltsev}, O., {Pavlov}, G.~G., \& {Romani}, R.~W. 2004, \apj, 602, 327

\bibitem[{{Kashyap} {et~al.}(2010){Kashyap}, {van Dyk}, {Connors}, {Freeman},
  {Siemiginowska}, {Xu}, \& {Zezas}}]{kashyap10}
{Kashyap}, V.~L., {van Dyk}, D.~A., {Connors}, A., {et~al.} 2010, \apj, 719,
  900

\bibitem[{{Klochkov} {et~al.}(2013){Klochkov}, {P{\"u}hlhofer}, {Suleimanov},
  {Simon}, {Werner}, \& {Santangelo}}]{klochkov13}
{Klochkov}, D., {P{\"u}hlhofer}, G., {Suleimanov}, V., {et~al.} 2013, \aap,
  556, A41

\bibitem[{{Lallement} {et~al.}(2014){Lallement}, {Vergely}, {Valette},
  {Puspitarini}, {Eyer}, \& {Casagrande}}]{lallement14}
{Lallement}, R., {Vergely}, J.-L., {Valette}, B., {et~al.} 2014, \aap, 561, A91

\bibitem[{{Larson} \& {Link}(1999)}]{larson99}
{Larson}, M.~B., \& {Link}, B. 1999, \apj, 521, 271

\bibitem[{{Lindegren} {et~al.}(2016){Lindegren}, {Lammers}, {Bastian},
  {Hern{\'a}ndez}, {Klioner}, {Hobbs}, {Bombrun}, {Michalik}, {Ramos-Lerate},
  {Butkevich}, {Comoretto}, {Joliet}, {Holl}, {Hutton}, {Parsons},
  {Steidelm{\"u}ller}, {Abbas}, {Altmann}, {Andrei}, {Anton}, {Bach},
  {Barache}, {Becciani}, {Berthier}, {Bianchi}, {Biermann}, {Bouquillon},
  {Bourda}, {Br{\"u}semeister}, {Bucciarelli}, {Busonero}, {Carlucci},
  {Casta{\~n}eda}, {Charlot}, {Clotet}, {Crosta}, {Davidson}, {de Felice},
  {Drimmel}, {Fabricius}, {Fienga}, {Figueras}, {Fraile}, {Gai}, {Garralda},
  {Geyer}, {Gonz{\'a}lez-Vidal}, {Guerra}, {Hambly}, {Hauser}, {Jordan},
  {Lattanzi}, {Lenhardt}, {Liao}, {L{\"o}ffler}, {McMillan}, {Mignard}, {Mora},
  {Morbidelli}, {Portell}, {Riva}, {Sarasso}, {Serraller}, {Siddiqui}, {Smart},
  {Spagna}, {Stampa}, {Steele}, {Taris}, {Torra}, {van Reeven}, {Vecchiato},
  {Zschocke}, {de Bruijne}, {Gracia}, {Raison}, {Lister}, {Marchant},
  {Messineo}, {Soffel}, {Osorio}, {de Torres}, \& {O'Mullane}}]{lindegren16}
{Lindegren}, L., {Lammers}, U., {Bastian}, U., {et~al.} 2016, \aap, 595, A4

\bibitem[{{Lutz} \& {Kelker}(1973)}]{lutz73}
{Lutz}, T.~E., \& {Kelker}, D.~H. 1973, \pasp, 85, 573

\bibitem[{{Martin} {et~al.}(2005){Martin}, {Fanson}, {Schiminovich},
  {Morrissey}, {Friedman}, {Barlow}, {Conrow}, {Grange}, {Jelinsky},
  {Milliard}, {Siegmund}, {Bianchi}, {Byun}, {Donas}, {Forster}, {Heckman},
  {Lee}, {Madore}, {Malina}, {Neff}, {Rich}, {Small}, {Surber}, {Szalay},
  {Welsh}, \& {Wyder}}]{martin05}
{Martin}, D.~C., {Fanson}, J., {Schiminovich}, D., {et~al.} 2005, \apjl, 619,
  L1

\bibitem[{{Mignani}(2011)}]{mignani11}
{Mignani}, R.~P. 2011, Advances in Space Research, 47, 1281

\bibitem[{{{\"O}zel} {et~al.}(2016){{\"O}zel}, {Psaltis}, {G{\"u}ver}, {Baym},
  {Heinke}, \& {Guillot}}]{ozel16a}
{{\"O}zel}, F., {Psaltis}, D., {G{\"u}ver}, T., {et~al.} 2016, \apj, 820, 28

\bibitem[{{Pavlov} {et~al.}(2017){Pavlov}, {Rangelov}, {Kargaltsev},
  {Reisenegger}, {Guillot}, \& {Reyes}}]{pavlov17}
{Pavlov}, G.~G., {Rangelov}, B., {Kargaltsev}, O., {et~al.} 2017, \apj, 850, 79

\bibitem[{{Petrovich} \& {Reisenegger}(2010)}]{petrovich10}
{Petrovich}, C., \& {Reisenegger}, A. 2010, \aap, 521, A77

\bibitem[{{Potekhin} {et~al.}(2015){Potekhin}, {Pons}, \& {Page}}]{potekhin15}
{Potekhin}, A.~Y., {Pons}, J.~A., \& {Page}, D. 2015, \ssr, 191, 239

\bibitem[{{Prakash} {et~al.}(1988){Prakash}, {Lattimer}, \&
  {Ainsworth}}]{prakash88}
{Prakash}, M., {Lattimer}, J.~M., \& {Ainsworth}, T.~L. 1988, Physical Review
  Letters, 61, 2518

\bibitem[{{Rajagopal} \& {Romani}(1996)}]{rajagopal96}
{Rajagopal}, M., \& {Romani}, R.~W. 1996, \apj, 461, 327

\bibitem[{{Rangelov} {et~al.}(2017){Rangelov}, {Pavlov}, {Kargaltsev},
  {Reisenegger}, {Guillot}, {van Kerkwijk}, \& {Reyes}}]{rangelov17}
{Rangelov}, B., {Pavlov}, G.~G., {Kargaltsev}, O., {et~al.} 2017, \apj, 835,
  264

\bibitem[{{Reisenegger}(1995)}]{reisenegger95}
{Reisenegger}, A. 1995, \apj, 442, 749

\bibitem[{{Schlafly} \& {Finkbeiner}(2011)}]{schlafly11}
{Schlafly}, E.~F., \& {Finkbeiner}, D.~P. 2011, \apj, 737, 103

\bibitem[{{Schlegel} {et~al.}(1998){Schlegel}, {Finkbeiner}, \&
  {Davis}}]{schlegel98}
{Schlegel}, D.~J., {Finkbeiner}, D.~P., \& {Davis}, M. 1998, \apj, 500, 525

\bibitem[{{Schultz} \& {Wiemer}(1975)}]{schultz75}
{Schultz}, G.~V., \& {Wiemer}, W. 1975, \aap, 43, 133

\bibitem[{{Shibanov} {et~al.}(1992){Shibanov}, {Zavlin}, {Pavlov}, \&
  {Ventura}}]{shibanov92}
{Shibanov}, I.~A., {Zavlin}, V.~E., {Pavlov}, G.~G., \& {Ventura}, J. 1992,
  \aap, 266, 313

\bibitem[{{Shibazaki} \& {Lamb}(1989)}]{shibazaki89}
{Shibazaki}, N., \& {Lamb}, F.~K. 1989, \apj, 346, 808

\bibitem[{{Shklovskii}(1970)}]{shklovskii70}
{Shklovskii}, I.~S. 1970, \sovast, 13, 562

\bibitem[{{Steiner} {et~al.}(2018){Steiner}, {Heinke}, {Bogdanov}, {Li}, {Ho},
  {Bahramian}, \& {Han}}]{steiner18}
{Steiner}, A.~W., {Heinke}, C.~O., {Bogdanov}, S., {et~al.} 2018, \mnras, 476,
  421

\bibitem[{{Tiengo} {et~al.}(2011){Tiengo}, {Mignani}, {de Luca}, {Esposito},
  {Pellizzoni}, \& {Mereghetti}}]{tiengo11}
{Tiengo}, A., {Mignani}, R.~P., {de Luca}, A., {et~al.} 2011, \mnras, 412, L73

\bibitem[{{Tody}(1993)}]{tody93}
{Tody}, D. 1993, in Astronomical Society of the Pacific Conference Series,
  Vol.~52, Astronomical Data Analysis Software and Systems II, ed. R.~J.
  {Hanisch}, R.~J.~V. {Brissenden}, \& J.~{Barnes}, 173

\bibitem[{{Verbiest} {et~al.}(2010){Verbiest}, {Lorimer}, \&
  {McLaughlin}}]{verbiest10}
{Verbiest}, J.~P.~W., {Lorimer}, D.~R., \& {McLaughlin}, M.~A. 2010, \mnras,
  405, 564

\bibitem[{{Yakovlev} \& {Pethick}(2004)}]{yakovlev04}
{Yakovlev}, D.~G., \& {Pethick}, C.~J. 2004, \araa, 42, 169

\bibitem[{{Young} {et~al.}(1999){Young}, {Manchester}, \& {Johnston}}]{young99}
{Young}, M.~D., {Manchester}, R.~N., \& {Johnston}, S. 1999, \nat, 400, 848

\bibitem[{{Zavlin} \& {Pavlov}(2004)}]{zavlin04}
{Zavlin}, V.~E., \& {Pavlov}, G.~G. 2004, \apj, 616, 452

\bibitem[{{Zavlin} {et~al.}(1996){Zavlin}, {Pavlov}, \& {Shibanov}}]{zavlin96}
{Zavlin}, V.~E., {Pavlov}, G.~G., \& {Shibanov}, Y.~A. 1996, \aap, 315, 141

\end{thebibliography}

\end{document}